\documentclass[acmsmall,screen]{acmart}

\AtBeginDocument{%
  }

\setcopyright{acmlicensed}
\copyrightyear{2026}
\acmYear{2026}
\acmDOI{XXXXXXX.XXXXXXX}
\acmConference[Conference]{Conference}{2026}{NY}
\acmISBN{978-1-4503-XXXX-X/2018/06}

\begin{document}
\author{Fangtian Zhong}
\author{Ollie Wold}
\author{Joseph Windmann}
\title{Detecting Call Graph Unsoundness without Ground Truth}


\renewcommand{\shortauthors}{Trovato et al.}
\newcommand{\parhead}[1]{\noindent\textbf{#1}}
\newcommand{\hdrstrut}{\rule{0pt}{2.4ex}}
\begin{abstract}
Java static analysis frameworks are commonly compared under the assumption that analysis algorithms and configurations compose monotonically and yield semantically comparable results across tools. In this work, we show that this assumption is fundamentally flawed. We present a large-scale empirical study of semantic consistency within and across four widely used Java static analysis frameworks: Soot, SootUp, WALA, and Doop. Using precision partial orders over analysis algorithms and configurations, we systematically identify violations where increased precision introduces new call-graph edges or amplifies inconsistencies. Our results reveal three key findings. First, algorithmic precision orders frequently break within frameworks due to modern language features such as lambdas, reflection, and native modeling. Second, configuration choices strongly interact with analysis algorithms, producing synergistic failures that exceed the effects of algorithm or configuration changes alone. Third, cross-framework comparisons expose irreconcilable semantic gaps, demonstrating that different frameworks operate over incompatible notions of call-graph ground truth.
These findings challenge prevailing evaluation practices in static analysis and highlight the need to reason jointly about algorithms, configurations, and framework semantics when assessing precision and soundness.

\end{abstract}

\ccsdesc[500]{Software and its engineering~Static analysis}
\ccsdesc[300]{Software and its engineering~Software testing}
\ccsdesc{Software and its engineering~Software reliability}

\keywords{Semantic Violations, Metamorphic Testing, Partial Order, Program analysis}

\maketitle

\section{Introduction}
Java remains a cornerstone of enterprise software and cloud infrastructure. Within these ecosystems, Java static analysis frameworks constitute critical infrastructure for ensuring software security, reliability, and performance \cite{lam2011soot, walagithub}. They underpin a wide range of downstream applications, including vulnerability detection \cite{zhang2022automatic, zhang2022example}, malware analysis \cite{xi2019deepintent}, and program optimizations \cite{vallee2000optimizing}. Consequently, errors in Java static analysis frameworks directly compromise the reliability of the downstream decisions that depend on them. Prior work has identified a variety of observable failures in these frameworks, such as crashes \cite{spoto2016julia}, memory leaks \cite{ghanavati2020memory,reif2019judge}, and concurrency bugs \cite{mansur2020detecting}. While these failures are important, they are typically self-revealing: analysis crashes, terminates abnormally, or fails to scale, making them comparatively easier to diagnose.

In contrast, semantic violations represent a more subtle and dangerous class of failures. A semantic violation occurs when an analysis completes successfully and produces plausible outputs that nonetheless violate expected semantic properties \cite{qian2025software}. Such failures are particularly harmful in security-critical contexts. For example, a single missing call-graph edge may cause an entire data-flow path to be missed, allowing vulnerabilities such as Log4Shell to evade detection \cite{crowdstrike_log4j}. As a result, semantic violations undermine trust in automated security auditing, refactoring, and maintenance tools. Detecting such violations is challenging for several reasons. First, they are fine-grained and easily obscured by large analysis outputs. Second, they often arise not from simple implementation errors but from subtle mismatches between theoretical models and practical implementations of complex language features. Third, for real-world programs, authoritative ground truth is generally unavailable; when different analyzers disagree, it is rarely clear which result is correct. Finally, although static analysis theory provides precision and soundness guarantees, these guarantees frequently break down in practice due to engineering trade-offs and incomplete language modeling.

To address both the absence of ground truth and the silent nature of semantic violations, we propose a methodology based on \emph{metamorphic testing} \cite{chen2018metamorphic}. Rather than validating analysis outputs against an external oracle, our approach checks internal semantic consistency within and across static analysis frameworks by exploiting expected relationships among analysis variants. Our key insight is that analysis algorithms and configuration settings are partially ordered by precision and soundness. For example, Class Hierarchy Analysis (CHA) \cite{dean1995optimization} must conservatively over-approximate Rapid Type Analysis (RTA) \cite{romano2018exploring}, which in turn must over-approximate Variable Type Analysis (VTA) \cite{crary1999flexible}. Any violation of this expected ordering, such as a more precise analysis missing true-positive call edges present in a less precise variant, or introducing edges where only pruning is expected, constitutes a semantic violation, independent of any external notion of correctness.

Guided by this principle, we conduct the first systematic intra-framework and cross-framework study of semantic consistency across four widely used Java static analysis frameworks. Our study spans multiple analysis algorithms, configuration settings, and their interactions. We uncover numerous previously undocumented semantic inconsistencies, many of which are triggered by modern Java language features introduced in JDK 8 and later. These findings reveal a substantial gap between theoretical expectations and practical implementations, particularly in the handling of contemporary language constructs inspired by functional programming.

We emphasize that deviations from theoretical precision or soundness orders do not necessarily imply specification-level errors. Instead, we treat partial orders as pragmatic semantic contracts that capture widely accepted expectations about the relative behavior of analysis algorithms and configurations. Violations of these contracts indicate semantic inconsistencies, regardless of whether their root causes stem from implementation defects, undocumented design decisions, or incomplete modeling of modern Java features. Our approach does not assume that such partial orders must hold universally; rather, violations serve as diagnostic signals that expose semantic mismatches between theory and implementation, which are particularly valuable to surface in the absence of ground truth and enable developers to distinguish between implementation defects, undocumented design decisions, and fundamental modeling limitations.  Our contributions include:

\begin{itemize}
    \item We introduce a ground-truth-free methodology for detecting semantic inconsistencies in Java static analysis frameworks by treating expected precision and soundness relationships as testable semantic contracts. This enables systematic testing in settings where labeled results are unavailable or impractical to obtain.
    
    \item We generalize partial-order reasoning by unifying analysis algorithms and configuration settings within and across a single framework. This allows us to expose semantic inconsistencies that arise from algorithm–configuration and inter-framework interactions, which are invisible to existing testing approaches.
    
    \item We conduct the first large-scale, cross-framework empirical study of semantic consistency across four widely used Java static analysis frameworks, uncovering previously undocumented violations and identifying recurring failure modes triggered by modern Java language features.
    
    \item how to assess their capabilities in practice
    \item How to determine the selection of algorithms and configurations when analyzing real applications
\end{itemize}

\section{Background and Motivation}
\label{sec:related}
This section motivates our work through a concrete example that highlights the challenges of testing modern Java static analysis frameworks. We then introduce the concept of partial orders over analysis algorithms and configuration settings, which we later extend to address these challenges.
\subsection{Motivating Example: The Lambda Resolution Failure} 
\label{subsec:motivating_example}
\begin{figure}[t]
  \centering
  \includegraphics[height=3in, width=\linewidth]{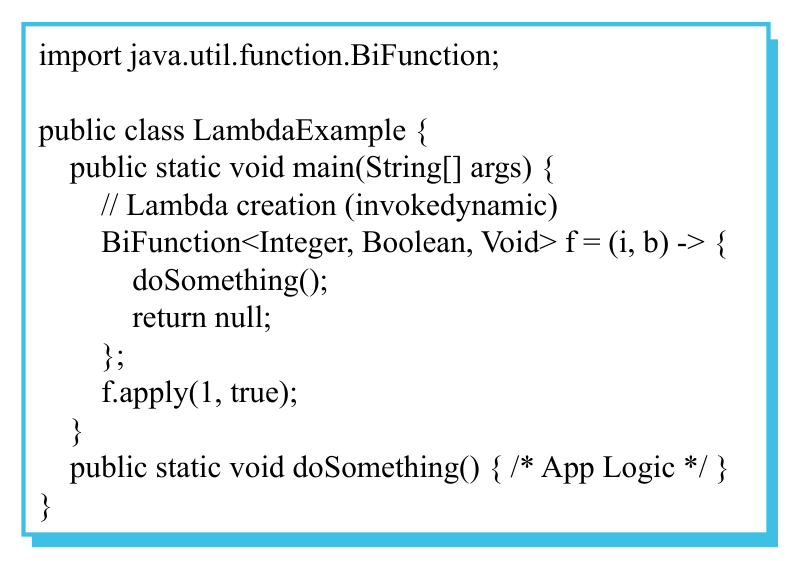}
  \caption{Example illustrating a lambda resolution failure in Soot.}
  \label{fig:lambda_failure}
\end{figure}

\begin{figure}[t]
  \centering
  \includegraphics[height=2.3in, width=\linewidth]{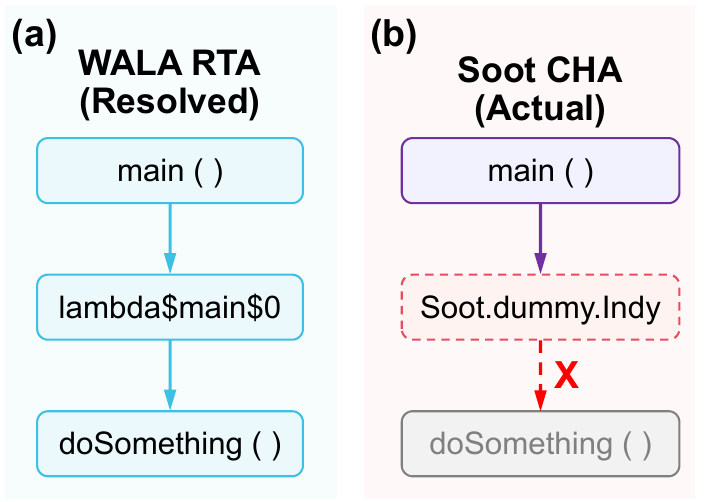}
  \caption{Example illustrating a lambda resolution failure in Soot IR.}
  \label{fig:IR}
\end{figure}
We begin with a simple yet representative example involving lambda expressions, which are widely used in modern Java programs. The example follows a common functional programming pattern in which a lambda expression is passed as a callback, as shown in Fig.~\ref{fig:lambda_failure}. From a developer’s point of view, the control flow is straightforward: the program creates an instance of a functional interface, which is later invoked to execute application-specific logic. A sound call graph should therefore resolve the invocation of the functional interface to the compiler-generated synthetic lambda method, and subsequently to the logic implemented inside the lambda.

However, as shown in Fig. \ref{fig:IR}, we observe inconsistent behavior across two widely used Java static analysis frameworks, WALA and Soot. Under their default configurations, WALA’s RTA correctly resolves the lambda metafactory and identifies the call to the application logic. In contrast, Soot’s CHA fails to resolve the lambda target and omits the corresponding call-graph edge entirely. As a result, the call to the application logic is missing from Soot’s generated call graph. This discrepancy arises from differences in how the two frameworks handle the JVM’s invokedynamic instruction. Java bytecode does not encode lambda targets as explicit method references; instead, the target is established at runtime via a bootstrap method \cite{yatish2019mining}. By default, Soot treats invokedynamic as an opaque call to a generic placeholder node. Because the connection to the synthetic lambda method (\emph{e.g.,} \emph{lambda\$main\$0}) is never established, the subsequent call to the application method \emph{doSomething()} is pruned from the call graph. WALA, by contrast, models \emph{LambdaMetafactory} semantics and resolves the lambda target. By missing the lambda edge, Soot incorrectly marks \emph{doSomething()} as unreachable, whereas WALA correctly treats it as reachable. As a result, any vulnerability located inside a lambda-wrapped callback becomes invisible to Soot-based analysis. For example, a taint analysis tracking data flows into \emph{doSomething()} would produce a false negative because the method appears unreachable in the call graph.

This example highlights a broader challenge in testing Java static analysis frameworks: semantic inconsistencies can emerge from interactions among analysis algorithms and configuration choices across different frameworks, yet remain difficult to detect in the absence of ground truth. Our work builds on the idea of reasoning about expected relationships between analysis results using partial orders. While Mordahl et al.~\cite{mordahl2023ecstatic} applied partial orders to configuration settings in Android taint analysis, we extend this notion to Java static analysis frameworks and generalize it across framework, algorithmic, and configuration dimensions. This generalization enables systematic detection of semantic inconsistencies even when labeled results are unavailable.

\parhead{Challenge 1: No Ground Truth} Constructing input programs with reliable ground truth is difficult when testing static analysis frameworks, especially for exposing semantic violations like the one described above in Soot. This difficulty is compounded by various analysis algorithms and configuration options. As program complexity increases, creating test cases with known expected behavior quickly becomes impractical. Without ground truth, disagreements between different analyses are inherently ambiguous, and determining which result is incorrect often requires significant manual effort and expertise.

\parhead{Challenge 2: Heterogeneous Analysis Algorithms and Configuration Options} Each framework provides its own set of analysis algorithms, and each algorithm exposes multiple configuration options. These choices vary across frameworks in availability, semantics and naming conventions. Testing algorithms or configurations in isolation can miss semantic bugs that arise from their interactions. To expose such bugs, it is necessary to test analysis algorithms and configuration options systematically, both within individual frameworks and across different frameworks.

\parhead{Challenge 3: Heterogeneous Intermediate Representations} 
 Static analysis frameworks operate over different intermediate representations. Soot and SootUp use Jimple \cite{lam2011soot, sootup_redesign}, WALA relies on SSA-based representations \cite{wala, ssa_representation}, and Doop expresses analysis declaratively using Datalog \cite{metamorphic_datalog}. Differences in representation design and call resolution strategies can lead to substantial naming differences in the resulting call graphs. Ongoing evolution of the Java language further amplifies these differences, as new language features are often supported unevenly across frameworks.
 
\subsection{Formal Partial Order Definitions}
\label{subsec:formaldefinitions}
To address the first challenge, our methodology adapts the concept of partial orders over analysis algorithms and configuration settings to reason about expected semantic relationships between analysis results. Related ideas have appeared in prior work, but at different levels of abstraction.
Grove~\cite{grove2001framework} introduced precision lattice relationships to characterize the relative precision of different static analysis algorithms, providing a theoretical ordering among analysis techniques.
More recently, Mordahl et al.~\cite{mordahl2021impact} applied partial orders over configuration settings to study how configuration choices influence the behavior of Android taint analysis tools.
Building on both lines of work, we unify algorithm-level and configuration-level partial-order reasoning.

\parhead{Precision Partial Order:} The key intuition behind a precision partial order is simple: a more precise analysis should conservatively refine a less precise one by eliminating false positives, not by introducing new ones. In other words, increasing precision should reduce spurious results while preserving all valid behavior. Formally, consider a set of programs and a static analysis algorithm that produces an output such as a call graph for each program. Each output contains edges that can be classified as either true positives or false positives. Let $P$ be the set of all programs. For a static analysis 
algorithm $A$, let $A(p)$ be the call graph it produces on input program $p$ where $p \in P$. $TP(\cdot)$ and $FP(\cdot)$ denote the sets of true positive and false positive edges, respectively. For two algorithms $A_1$ and $A_2$, it says that $A_1$ is \emph{at least as precise as} $A_2$, written $A_1 \succeq_r A_2$, iff $\forall p \in P,  FP(A_1(p)) \subseteq FP(A_2(p)).$ Intuitively, $A_1$ introduces no more spurious edges than $A_2$. Accordingly, a precision error occurs if
$A_1 \succeq_r A_2 \land (FP(A_1(p)) - FP(A_2(p)) \neq \emptyset)$. The same reasoning applies to configuration setting within a single analysis algorithm. Given two configuration setting, \( C_1 \) and \( C_2 \), we say that $C_1$ is \emph{at least as precise as} $C_2$, written $C_1 \succeq_r C_2$, iff $\forall p \in P,  FP(C_1(p)) \subseteq FP(C_2(p)).$ Accordingly, a precision error occurs if
$C_1 \succeq_r C_2 \land (FP(C_1(p)) - FP(C_2(p)) \neq \emptyset)$. 

\parhead{Soundness Partial Order:} The definition for a sound partial order would be: a more sound analysis should not refine a less sound one by eliminating feasible edges. Let \( A \) represent a algorithm, for two algorithms \( A_1 \) and \( A_2 \), saying that \( A_1 \) is at least as sound as \( A_2 \)), written $A_1 \succeq_s A_2$, iff $\forall p \in P,  TP(A_1(p)) \supseteq TP(A_2(p)).$ Intuitively, $A_1$ introduces more true edges than $A_2$. Accordingly, a soundness error occurs if
$A_1 \succeq_s A_2 \land (TP(A_2(p)) - TP(A_1(p)) \neq \emptyset)$. Similarly, for two configuration settings of an analysis algorithm, \( C_1 \) and \( C_2 \), it says that $C_1$ is \emph{at least as sound as} $C_2$, written $C_1 \succeq_s C_2$, iff $\forall p \in P,  TP(C_1(p)) \supseteq TP(C_2(p)).$ Accordingly, a soundness error occurs if
$C_1 \succeq_s C_2 \land (TP(C_2(p)) - TP(C_1(p)) \neq \emptyset)$. 

\section{Related Works} 

\subsection{Configuration Oriented Studies}
A large body of prior work has examined how configuration choices affect the behavior and effectiveness of static analysis tools. In the context of Android taint analysis, Pauck \emph{et al.} evaluated whether multiple tools satisfy the lifecycle, callback, inter-component communication guarantees of their default configurations and introduced ReproDroid to enable unified execution and result comparison \cite{pauck2018android}. Qiu \emph{et al.} systematically studied configuration options in FlowDroid, DroidSafe, and Amandroid, tuning them to support comparable language features before performing cross tool evaluation \cite{qiu2018analyzing}. Boxler compared FlowDroid, IccTA, and DroidSafe to assess analysis quality under fixed configurations \cite{boxler2018static}. Similar trends appear in studies of configurable Java static analysis frameworks. Smaragdakis \emph{et al.} modeled the design space of object sensitive analyses and evaluated how different sensitivity choices affect precision and performance \cite{smaragdakis2011pick}.  Wei \emph{et al.} evaluated a Java numeric analysis built on WALA by exploring different configurations and demonstrated complex tradeoffs between numeric and heap abstractions \cite{wei2018evaluating}. Configuration exploration has also been studied from a software engineering perspective. Configuration fuzzing techniques such as ConfigFuzz use coverage feedback to explore configuration spaces \cite{zhang2022registered}, while POWER employs staged exploration to uncover influential configuration combinations \cite{lee2022power}. Software product line research modeled configuration impact on performance using approaches such as CART \cite{guo2013variability}, DECART \cite{siegmund2015performance}, SPL Conqueror \cite{siegmund2012spl}, and DeepPerf \cite{ha2019deepperf}.

These studies provide useful insights into configuration sensitivity, but they typically evaluate tools under single configurations and do not examine how combinations of configurations affect analysis results. Even when large configuration spaces are explored, interactions between analysis algorithms and configurations are not considered. Such studies generally focus on configuration exploration within individual frameworks rather than cross framework semantic consistency. Moreover, much of the prior work primarily targets performance or other non functional properties rather than semantic correctness. A major limitation of these studies is their reliance on explicit ground truth, which prevents automatic large-scale analysis.

\subsection{Analysis Algorithm Studies}
A separate line of research focuses on the correctness and precision of analysis algorithms themselves. Lhoták and Hendren \cite{lhotak2003scaling} empirically evaluated context-sensitive points-to and call graph analyses within Soot, comparing precision tradeoffs across algorithm variants \cite{lhotak2008evaluating}. Samhi \emph{et al.} systematically quantify soundness gaps in Android static analysis tools by comparing statically inferred behaviors against dynamically observed executions, identifying implicit lifecycle callbacks and framework interactions as major sources of unsoundness \cite{samhi2024call}. Smaragdakis \emph{et al.} investigated tradeoffs among context sensitive call graph algorithms, highlighting how algorithmic choices affect precision \cite{precision_guided}. Reif \emph{et al.} introduced CATS to evaluate unsoundness in call graph construction algorithms, demonstrating that widely used algorithms may miss feasible call edges under certain language features \cite{reif2019judge}.
Other empirical studies have compared call graph algorithms within individual frameworks to characterize differences in precision and scalability \cite{zhang2024characterizing,sirlanci2025empirical,pang2021sok}. 

While these studies provide valuable insights into algorithmic behavior, they typically evaluate algorithms under fixed or implicit configurations and do not ensure that configuration settings are semantically equivalent across frameworks. As a result, they implicitly assume that configuration choices do not materially affect algorithmic relationships, an assumption that can confound semantic inconsistencies arising from algorithm–configuration interactions. Furthermore, most of these approaches rely on some form of ground truth, such as dynamic traces, handcrafted benchmarks, or reference implementations, which limits their applicability in real-world settings where authoritative ground truth is unavailable. The problem of analyzing cross-framework semantic consistency among  analysis algorithms, without relying on ground truth, also remains largely unexplored.

\subsection{Summary}
Overall, existing studies treat configuration testing and algorithm evaluation as largely independent concerns. They neither analyze how configuration choices reshape algorithmic behavior nor examine the semantic consequences of algorithm–configuration interactions across frameworks. This separation limits their ability to detect systematic semantic inconsistencies in static analysis results within/across static analysis frameworks, especially in the absence of ground truth.
  \begin{figure*}[t]
  \centering
  \includegraphics[ width=\linewidth]{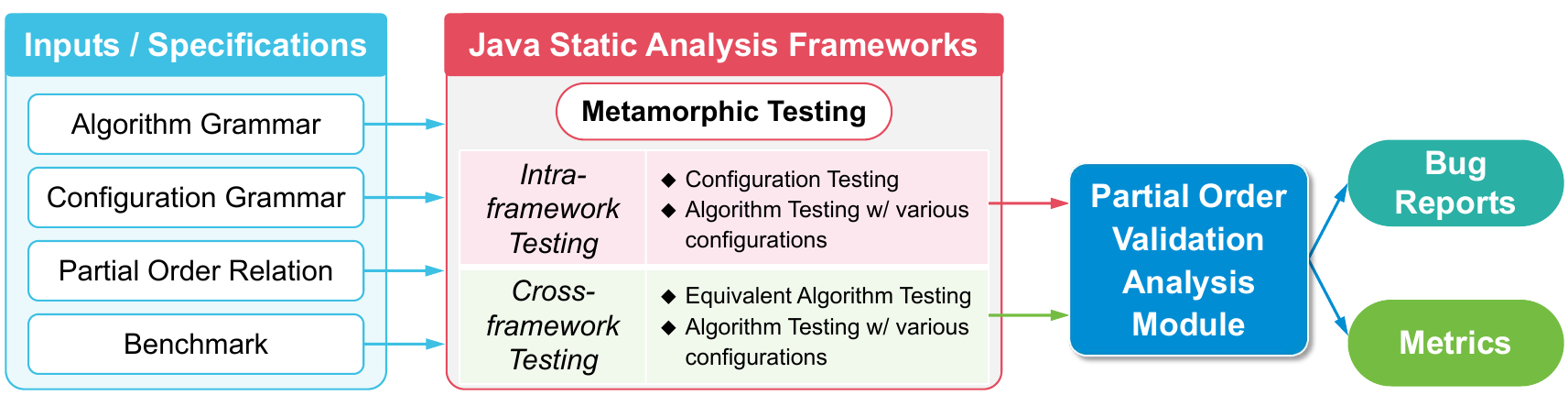}
  \caption{Overview of configuration and algorithm aware testing and debugging for static analysis}
  \label{fig:methodology}
\end{figure*}

\section{Partial Order Aware Testing}
\label{sec:design}
We propose a holistic approach to improving the automation of testing Java static analysis frameworks by leveraging partial orders over analysis algorithms and configuration settings. An overview of the approach is shown in Fig. \ref{fig:methodology}. Our method relies only on inputs that are realistic for framework developers and advanced users to obtain. First, it requires an algorithm grammar and a configuration grammar that formally describe the analysis algorithms supported by a framework and the configuration options available for each algorithm. Together, these grammars define the valid combinations of algorithms and configurations that the framework can execute. Second, it requires a specification of partial orders over algorithms, configurations, and algorithm–configuration pairs. These partial orders act as an internal semantic oracle that captures the expected relationships among different analysis choices. Formalizing these relationships enables systematic automated testing and  improves transparency by making a framework’s semantic guarantees explicit. Third, the approach requires a set of benchmark input programs without ground truth. 

Using these inputs, we construct a unified testing pipeline that supports two complementary testing dimensions. Within a single framework, we perform configuration testing, algorithm testing, and algorithm–configuration pair testing to detect intra-framework semantic inconsistencies. Across frameworks, we perform cross-framework testing, including equivalent-algorithm testing, algorithm comparisons, and algorithm–configuration pair testing, to identify semantic divergence in how identical bytecode is interpreted. All testing phases are driven by partial-order-aware metamorphic testing, which allows violations to be detected without ground truth. All analysis results are then processed by a centralized partial order validation module. This module checks observed behaviors against the specified partial orders and reports any violations as semantic inconsistency, which are aggregated into structured reports, and the framework also produces quantitative metrics summarizing the frequency, types, and distribution of partial order violations. These metrics provide actionable feedback for framework developers and maintainers.

\subsection{Improving Precision Partial Order Definitions}
\label{subsec:improveddefinitions}

The traditional definition of precision partial orders has an important limitation: it requires access to ground truth in order to classify analysis results and detect violations. This requirement is often infeasible, especially for large programs or complex analysis where ground truth is unavailable or prohibitively expensive to construct. To address these limitations, we introduce two improvements that extend precision partial orders to enable violation detection without relying on ground truth, substantially improving the automation of testing.

\parhead{Improvement 1: Implicit Soundness from Precision.} Existing work defines precision partial orders explicitly, either over analysis algorithms or over configuration settings. We extend this idea by introducing implicit soundness partial orders that are induced by precision partial orders. Consider two analysis algorithms, $A_1$ and $A_2$ where $A_1 \succeq_r A_2$. Precision ordering implies that $A_1$ should not introduce more false positives than $A_2$. Beyond this, we observe that increasing precision is generally not expected to change which behaviors are truly possible in the program. In other words, the set of true positives should remain stable. Based on this observation, we derive an implicit soundness relationship between the two algorithms $A_1 \succeq_S A_2$ and  $A_2 \succeq_S A_1$.

\parhead{Improvement 2: Analysis Algorithm Partial Orders with Configuration Settings.}
Precision partial orders are commonly defined separately for analysis algorithms and for configuration options. We extend this reasoning to algorithm–configuration pairs and introduce implicit soundness relations at this combined level. When considering algorithm--configuration pairs, if both the algorithm and the configuration follow their respective precision partial orders, the induced soundness relation remains unchanged.
That is, for $A_1 \succeq_r A_2$ and $C_1 \succeq_r C_2$, the pair $(A_1, C_1)$ should be more precise than $(A_2, C_2)$. Beyond expecting $(A_1, C_1)$ never produces more false positives than $(A_2, C_2)$, we also expect changing precision is generally not expected to alter the set of true positives. Thus, for the above precision partial order, we additionally add implicit soundness partial order $(A_1, C1_)\succeq_S (A_2, C_2)$ and  $(A_2, C_2) \succeq_S (A_1, C_1)$. This allows us to reason about semantic consistency across combinations of algorithms and configuration settings, rather than treating them in isolation.

\parhead{Detection Without Ground Truth.} Finally, we show how these improvements enable the detection of precision violations without relying on ground truth. We illustrate the idea using algorithm–configuration precision violations. Consider two pairs of algorithm-configuration, $(A_1, C_1)$ and $(A_2, C_2)$, applied to the same input program $p$. If $A_1(C_1, p) \succeq_r A_2(C_2, p)$, two conditions need to be satisfied for partial orders $A_1(C_1, p) \succeq_r A_2(C_2, p)$ and $A_2(C_2, p) \succeq_s A_1(C_1, p)$. Then, without ground truths, we can ascertain:
\begin{equation}
\begin{aligned}
TP(A_1(C_1, p)) &\subseteq TP(A_2(C_2, p)) 
&\lor FP(A_1(C_1, p)) \subseteq FP(A_2(C_2, p))
\end{aligned}
\end{equation}

 Although we cannot pinpoint the violation to one partial order, if both are defined, we can say $A_1(C_1, p) \not\succeq_r A_2(C_2, p)$ violates either $A_1(C_1, p) \succeq_r A_2(C_2, p)$ or $A_2(C_2, p) \succeq_s A_1(C_1, p)$. That is to say, if there is an edge that exists in $(A_1, C_1)$ but not in $A_2(C_2, p)$, this is considered as a algorithm-configuration precision partial order violation. While this method is less precise than approaches that rely on ground truth, it enables the detection of erroneous behavior in settings where ground truth is unavailable. 

 \subsection{Metamorphic Intra- and Inter- Framework Testing}
To address the second challenge, we propose a two-dimensional testing approach that combines intra-framework testing and cross-framework testing. The approach constructs partially ordered configurations, analysis algorithms, and algorithm–configuration pairs, executes them on input programs, and detects semantic violations by checking whether expected partial-order relationships are preserved. First, we perform intra-framework configuration testing based on configuration partial orders. For each framework, we define a baseline configuration using the framework’s context insensitive settings. For example, in Soot, we evaluate the core call-graph algorithms under their default configurations. Starting from this baseline, we progressively enable additional features such as field sensitivity, object sensitivity, and their combinations. These configuration changes have clear counterparts across frameworks (e.g., type-based and field-based options in Soot correspond to object-sensitive and field-sensitive options in WALA). After baseline testing, we exhaustively enumerate all valid combinations of supported configuration options within each framework. This allows us to systematically generate partially ordered configuration variants and uncover semantic violations caused by interactions among multiple non-baseline settings. Each configuration setting is applied to all programs, producing a comprehensive set of analysis results for comparison. Second, we perform intra-framework algorithm testing. Analysis algorithms within the same framework are evaluated under identical configurations as well as under different configurations, following partial orders derived from their theoretical algorithm relationships and configuration relationships. This testing dimension exposes semantic violations caused by algorithm refinement alone, as well as violations that arise from subtle interactions between algorithms and configurations. Third, for cross-framework comparisons, we apply partial order-aware testing to algorithm–configuration pairs whose theoretical relationships are well defined across frameworks. Additionally, we perform differential analysis between equivalent algorithms and configurations. This combination allows us to identify semantic divergence in how different frameworks interpret identical Java bytecode, even when they nominally implement the same analysis strategy.

\subsection{Partial Order Validation Analysis}

To address the third challenge and enable meaningful and fair quantitative comparison, we implemented a strict normalization pipeline for the outputs of Java static analysis frameworks. Raw analysis results often contain non-deterministic artifacts, such as memory addresses embedded in method signatures (for example, Demo.main\_11a00961) and randomized internal node identifiers. These artifacts vary across executions and tools and would otherwise dominate any direct comparison.
Before comparison, we normalize each call graph through the following steps. First, we remove all memory addresses and hash codes from method labels. Second, we resolve internal node identifiers to their corresponding semantic method signatures. Third, we filter out framework-specific metadata nodes that do not represent executable code, such as the “Cluster” nodes produced by Doop. Finally, we convert all method signatures into a unified \texttt{package.Class.method(args)} format to ensure compatibility across frameworks. This normalization step is essential. Without it, superficial syntactic differences between frameworks would lead to nearly zero similarity scores, even when the underlying call graphs represent the same program behavior. After graph normalization, we compare each pair of results on the same input program from two partially-ordered configurations, analysis algorithms, or algorithm-configuration pairs to detect violations within/across static analysis frameworks.  

\section{Evaluation}
\label{sec:evaluation}
\subsection{Experimental Setup}
\label{subsec:setup}
\parhead{Research questions:} Our empirical study aims to answer the following research questions:

\parhead{RQ1:} To what extent do configuration settings violate expected  partial orders within individual Java static analysis frameworks? RQ1 targets intra-framework configuration consistency.
It directly corresponds to our exhaustive enumeration of configuration combinations and evaluates whether enabling precision-related configuration options preserves expected partial-order relations.

\parhead{RQ2:} Do analysis algorithms within the same framework respect their theoretical precision partial orders under the same configurations? RQ2 focuses on intra-framework algorithmic consistency and isolates bugs caused by algorithm refinement alone.

\parhead{RQ3:} How do interactions between analysis algorithms and configuration choices amplify semantic violations within frameworks? RQ3 explicitly elevates algorithm–configuration pairs as objects of study. It justifies why enumerating all combinations is necessary and shows that configuration changes can  increase violations.

\parhead{RQ4:} How and why do Java static analysis frameworks diverge semantically when analyzing identical bytecode under partially ordered algorithm-configuration pairs and theoretically equivalent algorithms and configurations? RQ4 captures cross-framework semantic inconsistency under partially ordered metamorphic testing and differential analysis.

\parhead{Frameworks\&Benchmark:} We evaluate our approach on four widely used Java static analysis frameworks: Soot 4.5.0, SootUp 2.0.0, WALA 1.6.12, and Doop 4.24.13. Our evaluation is conducted using the JCG benchmark \cite{jcg2024_alt}, which has more than 125 Java programs and covers a broad spectrum of Java language features and program structures, including polymorphism, reflection, inheritance hierarchies, and generic collections. Besides, we use Jaccard Index to calculate the similarity between analysis results.

\subsection{RQ1: Intra-Framework Configuration Violations}
 We denote the baseline configuration as \emph{Baseline (BS)}, 
field-sensitive analysis as \emph{Field-Sensitive (FS)}, 
and object-sensitive analysis as \emph{Object-Sensitive (OS)}.
Following established results in pointer and call graph analysis, FS and OS are widely recognized as orthogonal precision dimensions that each strictly refine baseline, context-insensitive analyses. Prior work \cite{pointer,lhotak2008evaluating,hind2001pointer} shows that enabling either dimension improves precision, and that combining orthogonal refinements yields a strictly finer abstraction. We therefore assume the following precision partial order among configurations:
\[
\text{FS+OS}
\succeq_r
\{\text{FS}, \text{OS}\}
\succeq_r
\text{BS}.
\]

Intuitively, combining field sensitivity with object sensitivity yields a strictly finer abstraction than enabling either dimension alone.
Since only Soot and WALA expose configurable precision options beyond their baseline, we restrict configuration-level violation analysis to them.

In Soot, across all analysis algorithms, comparisons between the baseline and advanced configurations (OS, FS, and OS+FS), as well as comparisons among individual advanced configurations and their combined counterpart, produce identical call graphs. This indicates that, for our benchmark, Soot’s implementations of OS and FS do not further refine the call graph beyond the baseline and therefore conservatively preserve precision relationships. In contrast, WALA exhibits substantial violations of the precision partial orders. Enabling OS dramatically reshapes the call graph across all four analysis algorithms, yielding an average similarity of only 39.1\% relative to BS and resulting in 516 configuration-level violations. These violations indicate that OS introduces a large number of new call-graph edges that are absent under BS. The root cause of the violations between BS and OS lies in WALA’s handling of reflection. Under BS, WALA conservatively ignores ambiguous \texttt{Class.forName} calls unless the string argument is constant. When OS is enabled, however, the propagation of allocation-site contexts activates WALA’s specialized reflection handlers, which attempt to infer potential reflective targets based on available types in the analysis scope. This speculative recovery improves recall but also introduces additional call edges that violate the expected precision refinement.

FS alone exhibits markedly different behavior. Across all algorithms, FS maintains over 94.3\% similarity with BS and triggers only a single violation under RTA. The sole violation between the BS and FS configurations stems from how container classes are abstracted. Under BS, fields of containers such as \texttt{ArrayList} are merged into a single abstract location, leading to early loss of information and preventing certain object flows from being propagated. Enabling FS preserves these distinctions, allowing data flow through container operations that were previously blocked and thereby exposing the call edge that BS fails to capture. Interestingly, enabling FS in addition to OS introduces no further structural changes beyond those already introduced by OS. Moreover, the combined configuration OS+FS results in fewer violations than OS alone, indicating that FS partially constrains the additional edges introduced by OS. A similar stabilizing effect is observed when OS+FS is compared against FS alone. While OS introduces speculative edges through aggressive reflection recovery, the addition of FS simultaneously prunes invalid paths by constraining pointer propagation. As a result, BS–OS+FS and FS–OS+FS configuration pairs trigger substantially fewer violations, highlighting the role of FS in mitigating the semantic instability introduced by OS.

\begin{table*}[t]
\caption{WALA Configuration Results.
Each entry reports \textit{Similarity (\%) / \#Violations}.}
\label{tab:inter_algo_same_cfg}
\centering
\small
\setlength{\tabcolsep}{4pt}
\renewcommand{\arraystretch}{1.1}

\begin{tabular}{lcccc}
\toprule
\textbf{Alg.} &
\textbf{BS-FS} &
\textbf{BS-OS} &
\textbf{BS-OS+FS} &
\textbf{FS-OS+FS} \\
\midrule
RTA              & 94.3\%/1 & 43.3\%/143 & 43.3\%/143 & 45.9\%/125 \\
0-CFA            & 100\%/0  & 41.4\%/99 & 42.6\%/92  & 40.3\%/90 \\
0-Con.-CFA  & 99.5\%/0 & 37.2\%/134 & 40.8\%/97  & 43.3\%/94 \\
0-1-CFA          & 100\%/0  & 34.6\%/140 & 37.6\%/133 & 40.2\%/117 \\

\bottomrule
\end{tabular}
\end{table*}

\subsection{RQ2: Intra-framework Algorithmic Violations Under The Same Configurations}
\begin{table*}[t]
\caption{Inter-algorithm results under identical configurations.
Each entry reports \textit{Similarity (\%) / \#Violations}.}
\label{tab:inter_algo_same_config}
\centering
\small
\setlength{\tabcolsep}{4pt}
\renewcommand{\arraystretch}{1.1}

\begin{tabular}{lcccc}
\toprule
\textbf{Alg. Pair} &
\textbf{BS} &
\textbf{FS} &
\textbf{OS} &
\textbf{OS+FS} \\
\midrule
\multicolumn{5}{c}{\textbf{Panel A: Soot}} \\
\midrule
CHA$\to$RTA & 69.5/105 & 69.8/105 & 69.5/105 & 70.1/98 \\
CHA$\to$VTA & 68.2/98  & 68.2/98  & 68.2/98  & 69.0/92 \\
RTA$\to$VTA & 91.6/228 & 91.6/228 & 91.6/228 & 95.4/6 \\
\midrule
\multicolumn{5}{c}{\textbf{Panel B: WALA}} \\
\midrule
RTA$\to$0-CFA & 87.4/8 & 90.7/10 & 93.0/10 & 92.6/10 \\
RTA$\to$0-Con.-CFA & 87.8/8 & 94.6/7 & 93.0/10 & 91.8/10 \\
RTA$\to$0-1-CFA & 87.0/10 & 94.6/7 & 93.8/8 & 92.7/10 \\
\midrule
0-CFA$\to$0-Con.-CFA & 100/0 & 99.5/0 & 100/0 & 100/0 \\
0-CFA$\to$0-1-CFA & 99.6/0 & 99.5/0 & 100/0 & 100/0 \\
0-Con.-CFA$\to$0-1-CFA & 99.6/0 & 100/0 & 100/0 & 100/0 \\
\midrule
\multicolumn{5}{c}{\textbf{Panel C: Doop (BS only)}} \\
\midrule
Context-Insensitive$\to$1-Call-Site-Sens.-Heap & 100/0 & -- & -- & -- \\
Context-Insensitive$\to$1-Object-Sens.-Heap & Fail & -- & -- & -- \\
Context-Insensitive$\to$1-Type-Sens.-Heap & 100/0 & -- & -- & -- \\
1-Call-Site-Sens.-Heap$\to$1-Type-Sens.-Heap & 99.3/110 & -- & -- & -- \\

1-Call-Site-Sens.-Heap$\to$1-Object-Sens.-Heap & Fail & -- & -- & -- \\
\bottomrule
\end{tabular}
\end{table*}

We validate whether each framework adheres to the expected theoretical partial orders among its supported analysis algorithms, as established in prior work~\cite{shivers1991control,lhotak2003scaling,bravenboer2009strictly}. 
Specifically, the expected precision orders are as follows:
\begin{itemize}
  \item \textbf{Soot:} $\text{VTA} \succeq_r \text{RTA} \succeq_r \text{CHA}$
  \item \textbf{WALA:} $\text{0-1-CFA} \succeq_r \text{0-Container-CFA} \succeq_r \text{0-CFA} \succeq_r \text{RTA}$
  \item \textbf{Doop:} $\{\text{1-Object-Sensitive-Heap}, \text{1-Type-Sensitive-Heap}\} \succeq_r \text{1-Call-Site-Sensitive-Heap} \succeq_r \text{Context-Insensitive}$
  \item \textbf{SootUp:} $\text{RTA} \succeq_r \text{CHA}$
\end{itemize}

In SootUp, CHA and RTA produce highly similar call graphs, with a similarity of 95.8\%, yet still trigger six algorithm-level violations. This result demonstrates that semantic inconsistencies can arise even when the overall call-graph structure appears largely unchanged. A closer inspection reveals that the transition from CHA to RTA in SootUp exhibits a soundness inversion: RTA resolves valid call edges that CHA fails to capture. The root cause lies in SootUp’s handling of \texttt{invokedynamic}. CHA relies primarily on the static class hierarchy and therefore ignores targets of dynamic invocation sites, including lambda expressions. In contrast, RTA performs reachability-driven analysis and actively resolves lambda bootstrap methods, computing their concrete targets and adding the corresponding call edges. As a result, RTA discovers reachable execution paths that the baseline analysis pessimistically omits.

Table~\ref{tab:inter_algo_same_config} Panel A shows that, under identical configurations, both RTA and VTA in Soot prune subsets of edges relative to CHA. In this setting, call-graph similarity between CHA and RTA or VTA remains relatively low, at approximately 68.2\%–69.8\%, indicating that switching from CHA to either refined algorithm substantially reshapes the call graph. By contrast, the transition from RTA to VTA results in fewer structural changes, as both analyses prune largely overlapping subsets of CHA edges. The remaining violations in this transition arise because VTA over-approximates variable types in certain flows and therefore resolves virtual calls that RTA conservatively prunes away. Finally, enabling \textbf{OS+FS} slightly increases call-graph similarity while dramatically reducing algorithm-level violations, particularly for the RTA-to-VTA transition, where violations drop from 228 to 6. The reason is that, in Soot, combining OS and FS stabilizes inter-algorithm relationships by constraining spurious flows while preserving necessary reachability, thereby reducing inconsistencies.

Table~\ref{tab:inter_algo_same_config} Panel B reports WALA’s inter-algorithm behavior. Comparing RTA against more precise context-sensitive analyses (0-CFA, 0-Container-CFA, and 0-1-CFA) reveals a consistent pattern: across all configurations, these algorithm pairs exhibit relatively high call-graph similarity (approximately 87\%–94.6\%) while still triggering a non-trivial number of violations (7–10 per pairing). Root-cause analysis shows that these violations stem from systematic modeling limitations in WALA’s RTA rather than implementation noise. Although RTA is theoretically less precise, it exhibits structural regressions relative to more precise algorithms by failing to resolve several classes of reachable behavior. Specifically, RTA omits call edges associated with static initialization order, exception control flow in re-thrown catch blocks, and intrinsic handling of \texttt{MethodHandle} operations. In contrast, the context-sensitive analyses correctly capture these JVM semantics through more complete control-flow construction and intrinsic simulation, leading to semantically richer call graphs. By comparison, transitions among context-sensitive analyses (0-CFA → 0-Container-CFA, 0-CFA → 0-1-CFA, and 0-Container-CFA → 0-1-CFA) exhibit near-perfect similarity (99.5\%–100\%) and no violations under all configurations. This indicates that once a context-sensitive baseline is established, WALA’s refinement hierarchy becomes internally consistent.

As shown in Table \ref{tab:inter_algo_same_config} Panel C, in Doop, comparisons between the context-insensitive analysis and 1-Call-Site-Sensitive-Heap or 1-Type-Sensitive-Heap produce identical call graphs with no violations, indicating strict adherence to the expected algorithmic partial order at this refinement level. In contrast, the transition from 1-Call-Site-Sensitive-Heap to 1-Type-Sensitive-Heap, while exhibiting high structural similarity, triggers a large number of violations (110), revealing substantial semantic inconsistencies. These violations stem from precision loss in the 1-Type-Sensitive-Heap abstraction. Specifically, 1-Call-Site-Sensitive-Heap distinguishes heap objects by allocation site, whereas the 1-Type abstraction merges all objects of the same class, such as collapsing all \texttt{ArrayList} instances into a single abstract object. In the violating cases, this merging causes data flows originating from one container instance to leak into another, introducing spurious call edges that are provably unreachable under 1-Call-Site-Sensitive-Heap. Notably, 1-Object-Sensitive-Heap fails to complete since state space  exceeds the solver's resource limits in Doop. 

\subsection{RQ3: Intra-framework Algorithmic Violation Under Different Configurations}
\begin{table}[t]
\caption{WALA and Soot inter-algorithm results under different configurations (\textit{Sim./Vio.}).}
\label{tab:wala_soot_inter_diff_cfg}
\centering
\small
\setlength{\tabcolsep}{4pt}
\renewcommand{\arraystretch}{1.1}

\begin{tabular}{lccccc}
\toprule
\textbf{Alg. Pair} &
\textbf{BS--FS} &
\textbf{BS--OS} &
\textbf{BS--OS+FS} &
\textbf{FS--OS+FS} &
\textbf{OS--OS+FS} \\
\midrule
\multicolumn{6}{c}{\textbf{Panel A: WALA}} \\
\midrule
RTA$\to$0-CFA      & 84.0/8   & 33.0/146 & 33.4/135 & 35.9/117 & 92.6/10 \\
RTA$\to$0-Container-CFA & 87.7/8   & 33.0/146 & 35.7/109 & 38.0/99  & 91.8/10 \\
RTA$\to$0-1-CFA    & 87.7/8   & 32.2/144 & 33.0/139 & 35.5/121 & 92.7/10 \\
\midrule
0-CFA$\to$0-Container-CFA   & 99.5/0  & 41.4/99  & 43.6/86  & 42.2/77  & 100/0 \\
0-CFA$\to$0-1-CFA      & 99.5/8  & 38.1/99  & 42.6/92  & 39.7/94  & 100/0 \\
0-Container-CFA$\to$0-1-CFA & 99.5/0  & 34.9/134 & 37.3/127 & 40.2/117 & 100/0 \\
\bottomrule
\multicolumn{6}{c}{\textbf{Panel B: Soot}} \\
\midrule
CHA$\to$RTA & 82.8/224 & 82.5/224 & 83.1/76 & 83.1/76 & 83.1/76 \\
CHA$\to$VTA & 81.9/218 & 81.9/218 & 82.4/70 & 82.4/70 & 82.4/70 \\
RTA$\to$VTA & 91.6/228 & 91.6/228 & 95.4/6  & 95.4/6  & 95.4/6  \\
\bottomrule
\end{tabular}
\end{table}
 Table \ref{tab:wala_soot_inter_diff_cfg} Panel A reports inter-algorithm behavior under different configurations in WALA. Comparing RTA to context-sensitive analysis (0-CFA, 0-Container-CFA, and 0-1-CFA), inter-algorithm consistency degrades substantially when configurations differ. Under BS→FS, similarity remains moderately high (84.0\%–87.7\%), but under BS→OS and BS→OS+FS it drops sharply to approximately 33.2\%–35.7\%, accompanied by a dramatic increase in violations (up to 146). This contrasts sharply with Table \ref{tab:inter_algo_same_config}, where the same algorithm pairs under identical configurations maintain high similarity and trigger only a small number of violations (7–10). Together, these results show that, in WALA, configuration changes significantly amplify inter-algorithm inconsistencies that are largely suppressed when configurations are fixed. A similar pattern appears for transitions among context-sensitive algorithms (0-CFA → 0-Container-CFA, 0-CFA → 0-1-CFA, and 0-Container-CFA → 0-1-CFA). Under different configurations, similarity drops markedly for BS→OS, BS→OS+FS, and FS→OS+FS (to approximately 34.9\%–43.6\%), and a large number of violations are introduced (up to 134). This indicates that configuration mismatches disrupt the otherwise well-behaved ordering among context-sensitive analysis. Notably, comparisons involving OS→OS+FS in Table \ref{tab:wala_soot_inter_diff_cfg} consistently recover high similarity (91.8\%–100\%) with zero or few violations across all algorithm pairs. This mirrors the behavior observed in Table \ref{tab:inter_algo_same_config} under OS+FS and highlights a stabilizing effect of combining OS and FS, even when algorithm changes are coupled with configuration changes. 
 
 Compared to testing under identical configurations, this algorithm–configuration hybrid testing exposes a distinct failure mode. In several cases, the number of violations exceeds the sum of those observed under algorithm-only or configuration-only testing. This effect reflects a form of synergistic failure, where configuration changes and algorithm refinement interact to amplify semantic inconsistencies. For example, enabling OS introduces speculative reflection edges. When combined with flow-sensitive analyses such as 0-CFA, data flow propagates through these speculative edges, reaching downstream methods that were unreachable under BS configuration. In this setting, configuration choices create additional semantic uncertainty, and algorithmic precision amplifies the resulting error beyond what either factor produces in isolation.

In Soot, Table \ref{tab:wala_soot_inter_diff_cfg} Panel B reports inter-algorithm behavior under different configurations. Across all algorithm pairs (CHA→RTA, CHA→VTA, and RTA→VTA), switching configurations while changing algorithms yields moderate-to-high similarity (approximately 81.9\%–95.4\%) but introduces a large number of violations, particularly for transitions originating from CHA. For example, CHA→RTA and CHA→VTA incur over 200 violations under BS→FS and BS→OS, indicating that combining algorithm changes with configuration changes substantially amplifies inconsistencies. When compared to Table \ref{tab:inter_algo_same_config}, a clear pattern emerges. Under identical configurations, CHA→RTA and CHA→VTA already exhibit relatively low similarity (around 68\%–70\%) with over 100 violations, reflecting structural differences introduced by algorithmic pruning alone. However, Table \ref{tab:wala_soot_inter_diff_cfg} shows that these inconsistencies become more severe when configurations also differ, as evidenced by the sharp increase in violation counts (e.g., 224 for CHA→RTA under BS→FS). This demonstrates that configuration changes exacerbate inter-algorithm instability beyond what is observed under fixed configurations. For RTA→VTA, both tables reveal a stabilizing effect when OS+FS is enabled. Comparisons involving FS→OS+FS or OS→OS+FS achieve the highest similarity (95.4\%) and the lowest violation counts. One contributing factor is that field sensitivity enables RTA to track \texttt{MethodHandle} objects through fields. Under CHA, field reads involving such handles are treated as unknown and call resolution stops. In contrast, field-sensitive RTA can trace the handle to a concrete method and resolve the call. As a result, the more precise configuration enables the refined algorithm to discover reachable behavior that the less precise configuration misses. This observation confirms that, for dynamic Java features, increased precision is often a prerequisite for soundness rather than merely a means of pruning spurious edges.

\subsection{RQ4: Cross-Framework Semantic Violation}

\subsubsection{Cross-framework comparisons between algorithm-configuration pairs}
\begin{table*}[t]
\caption{SootUp $\to$ Soot across algorithms. Similarity confirms partial semantic overlap.}
\label{tab:sootupsootinteralg}
\centering
\small
\setlength{\tabcolsep}{4pt}
\renewcommand{\arraystretch}{1.1}

\begin{tabular}{lcccc}
\toprule
\textbf{Alg. Pair} &
\textbf{BS} &
\textbf{FS} &
\textbf{OS} &
\textbf{FS+OS} \\
\midrule
CHA $\to$ RTA & 32.2\%/270 & 32.2\%/270 & 32.2\%/270 & 32.2\%/270 \\
CHA $\to$ VTA & 31.8\%/259 & 31.8\%/259 & 31.8\%/259 & 31.8\%/259 \\
RTA $\to$ VTA & 31.7\%/416 & 31.7\%/416 & 31.7\%/416 & 31.7\%/416 \\
\bottomrule
\end{tabular}
\end{table*}

\begin{table*}[t]
\caption{Soot vs. WALA. Low similarity reflects fundamental disagreements on call graph ground truth between frameworks.}
\label{tab:soot_wala_compact_matrix}
\centering
\scriptsize
\setlength{\tabcolsep}{2.2pt}
\renewcommand{\arraystretch}{1.05}

\resizebox{\textwidth}{!}{%
\begin{tabular}{lccccccccc}
\toprule
\textbf{Alg. Pair} &
\hdrstrut{\textbf{BS--BS}} &
\hdrstrut{\textbf{BS--FS}} &
\hdrstrut{\textbf{BS--OS}} &
\hdrstrut{\textbf{BS--FS+OS}} &
\hdrstrut{\textbf{FS--FS}} &
\hdrstrut{\textbf{FS--FS+OS}} &
\hdrstrut{\textbf{OS--OS}} &
\hdrstrut{\textbf{OS--FS+OS}} &
\hdrstrut{\textbf{FS+OS--FS+OS}} \\
\midrule
CHA$\to$RTA &21.2\%/198 & 21.6\%/188 & 21.2\%/184 & 21.2\%/184 & 21.6\%/188 & 21.2\%/184 & 21.2\%/184 & 21.2\%/184 & 21.2\%/184 \\
CHA$\to$0-CFA &
12.0\%/138 & 11.1\%/141 & 10.0\%/185 & 10.5\%/173 & 11.1\%/141 & 10.5\%/173 & 10.0\%/185 & 10.5\%/173 & 10.5\%/173 \\
CHA$\to$0-Con. &
10.6\%/174 & 11.4\%/160 & 10.0\%/185 & 11.7\%/144 & 11.4\%/160 & 11.7\%/144 & 10.0\%/185 & 11.7\%/144 & 11.7\%/144 \\
CHA$\to$0-1-CFA &
10.0\%/184 & 11.4\%/160 & 10.5\%/177 & 10.3\%/177 & 11.4\%/160 & 10.3\%/177 & 10.5\%/177 & 10.3\%/177 & 10.3\%/177 \\
\midrule
RTA$\to$0-CFA &
12.0\%/138 & 11.1\%/141 & 10.0\%/185 & 10.5\%/173 & 11.1\%/141 & 10.5\%/173 & 10.0\%/185 & 10.5\%/173 & 10.5\%/173 \\
RTA$\to$0-Con. &
10.6\%/174 & 11.4\%/160 & 10.0\%/185 & 11.7\%/144 & 11.4\%/160 & 11.7\%/144 & 10.0\%/185 & 11.7\%/144 & 11.7\%/144 \\
RTA$\to$0-1-CFA &
10.0\%/184 & 11.4\%/160 & 10.5\%/177 & 10.3\%/177 & 11.4\%/160 & 10.3\%/177 & 10.5\%/177 & 10.3\%/177 & 10.3\%/177 \\
\midrule
VTA$\to$0-CFA &
12.0\%/138 & 11.1\%/141 & 10.0\%/185 & 10.5\%/173 & 11.1\%/141 & 10.5\%/173 & 10.0\%/185 & 10.5\%/173 & 10.5\%/173 \\
VTA$\to$0-Con. &
10.6\%/174 & 11.4\%/160 & 10.0\%/185 & 11.7\%/144 & 11.1\%/160 & 11.7\%/144 & 10.1\%/184 & 11.7\%/144 & 11.7\%/144 \\
VTA$\to$0-1-CFA &
10.0\%/184 & 11.4\%/160 & 10.5\%/177 & 10.3\%/177 & 11.4\%/160 & 10.3\%/177 & 10.5\%/177 & 10.3\%/177 & 10.3\%/177 \\
\bottomrule
\end{tabular}%
}
\end{table*}

We validated whether the expected theoretical precision partial order among analysis algorithms is preserved across different Java static analysis frameworks:
$\text{0-1-CFA} \succeq_r \text{0-Container-CFA} \succeq_r \text{0-CFA} \succeq_r \text{VTA} \succeq_r \text{RTA} \succeq_r \text{CHA}$. Because there is no proof establishing a precise correspondence between algorithms in Doop and other frameworks, we empirically compare Doop with other frameworks.

The comparison Soot (CHA) → SootUp (RTA) exhibits a moderate semantic overlap, with 33.0\% similarity and 240 violations, more violations than observed in any intra-framework comparison. Despite their shared lineage, this high violation count indicates that differences in semantic modeling and algorithm interpretation, rather than implementation noise, play a dominant role. Discrepancies arise from a soundness inversion: SootUp’s RTA implementation correctly resolves certain virtual calls that Soot prunes due to overly rigid type-based filtering. Besides, comparison between SootUp's less precise analysis algorithms and Soot's more precise analysis algorithms, Table~\ref{tab:sootupsootinteralg} shows that similarity for SootUp → Soot remains consistently around 32\% across all algorithm pairs, regardless of the configuration used in Soot. This invariance across BS–FS, BS–OS, and BS–FS+OS indicates that configuration changes in Soot have little effect on the cross-framework relationship. The root cause of these violations lies in SootUp’s handling of lambda expressions. SootUp’s CHA frequently truncates call graphs at lambda call sites by replacing \texttt{invokedynamic} instructions with dummy sink nodes, preventing resolution of lambda targets and causing it to miss valid execution paths that Soot’s RTA and VTA capture. When comparing SootUp RTA against Soot VTA, the number of violations increases further, reaching 416. This occurs because SootUp treats \texttt{invokedynamic} as an explicit sink (\texttt{dummy.InvokeDynamic}), effectively severing the call graph at lambda boundaries, whereas Legacy Soot’s VTA employs a more mature resolution pipeline that links these sites to their synthetic implementations. As a result, VTA introduces valid call edges to lambda bodies that are absent from the SootUp RTA graph, triggering the observed violations.

Table~\ref{tab:soot_wala_compact_matrix} presents a comprehensive cross framework comparison between Soot and WALA, in which call graphs produced by Soot’s CHA, RTA, and VTA are compared against WALA’s RTA, 0-CFA, 0-Container-CFA, and 0-1-CFA under multiple configurations. Across all algorithm pairs and configurations, call graph similarity remains consistently low, ranging from approximately 10\% to 21\%. Notably, this low similarity persists regardless of whether the comparison involves Soot’s least precise algorithm, CHA, or its more refined variants, RTA and VTA, indicating that increasing algorithmic precision in Soot does not reduce the semantic gap with WALA. Moreover, similarity remains largely invariant across configuration changes, including BS, FS, OS, and their combinations, suggesting that the observed divergence is not driven by configuration mismatches. The consistently high number of violations, typically between 138 and 198, further confirms that the disagreement is systematic rather than incidental. Taken together, these results reveal a fundamental design philosophy gap between Soot and WALA that arises from asymmetric semantic modeling choices rather than isolated implementation defects. A primary source of this divergence is reflection handling. WALA incorporates built in heuristics to resolve reflective targets, such as analyzing strings passed to \texttt{Class.forName}, enabling it to recover valid execution paths through reflective calls. In contrast, Soot treats reflection as opaque and therefore omits these execution paths entirely. 
When it comes to the inter framework comparison between WALA and Soot for the RTA and VTA algorithm pair. Across all configurations, similarity remains consistently low, centered around 21\%, with only minor fluctuations. Correspondingly, the number of violations remains uniformly high, ranging from 257 to 259. Discrepancies are primarily driven by entry point selection. Soot adopts a conservative reachability assumption that treats a broader set of methods, including library code, as potential entry points. WALA, by contrast, defaults to a stricter main class centric entry model and intentionally prunes library paths that are never invoked. As a result, Soot includes many edges corresponding to dead or unreachable code that WALA deliberately excludes.

Tables~\ref{tab:sootup_wala} examine the semantic relationship between SootUp and WALA. It reveals substantially low similarity values, ranging from approximately 10.1\% to 21\% across all algorithm pairs and WALA configurations. The highest similarity is observed for CHA → RTA (about 21\%), whereas comparisons involving WALA’s more context-sensitive analyses consistently yield lower similarity (approximately 10.1–11.9\%), accompanied by a large number of violations. The comparison between SootUp and WALA exposes a substantial maturity gap in their handling of contemporary Java bytecode features, reflected in an extremely low mean similarity of 3.9\%. Discrepancies are dominated by lambda handling. WALA’s SSA-based intermediate representation explicitly resolves invokedynamic bootstrap methods and successfully links lambda call sites to their concrete implementations. In contrast, SootUp currently truncates call graphs at invokedynamic sites, causing it to miss a large number of valid execution paths in programs that rely on Java 8+ features. As a result, WALA consistently outperforms SootUp on lambda-heavy benchmarks.  In the opposite direction, discrepancies arise primarily from differences in type-hierarchy modeling. SootUp relies on a cached class hierarchy and may retain call edges based solely on declared types, even when those edges are unreachable under WALA’s more precise, flow-sensitive pointer analysis. This divergence is further amplified by SootUp’s broader inclusion of phantom types, whereas WALA enforces a stricter analysis scope.

Comparisons between Doop and other frameworks yield similarity scores close to 0\%. It is important to clarify that such values do not indicate disjoint call graphs. Instead, they reflect a magnitude mismatch inherent in the similarity metric, which we compute using the Jaccard index:
\begin{equation}
J(A, B) = \frac{|A \cap B|}{|A \cup B|}
\end{equation}
Doop explicitly models implicit native and runtime flows, producing substantially larger call graphs ($|B| \approx 15,000$ edges) than the bytecode-constrained graphs generated by Soot or WALA ($|A| \approx 150$ edges). 
Even in the ideal case where Doop subsumes all edges identified by Soot ($A \subset B$), the union $|A \cup B|$ is dominated by Doop’s additional edges, yielding a similarity score of $J \approx = 0.01$ approaches zero. Therefore, near-zero similarity values primarily reflect differences in analysis scope rather than an inability to resolve application-level logic.

\begin{table}[t]
\caption{SootUp $\rightarrow$ WALA across algorithms and WALA configurations.
Each entry is \textit{Sim./Vio.}.}
\label{tab:sootup_wala}
\centering
\scriptsize
\setlength{\tabcolsep}{4pt}
\renewcommand{\arraystretch}{1.05}
\begin{tabular}{lcccc}
\toprule
\textbf{Alg. Pair} &
\textbf{BS} &
\textbf{FS} &
\textbf{OS} &
\textbf{FS+OS} \\
\midrule
CHA$\rightarrow$RTA        & 20.8\%/198 & 21.0\%/161 & 20.8\%/184 & 20.8\%/184 \\CHA$\rightarrow$0-CFA      & 11.9\%/138 & 11.0\%/141 & 10.2\%/185 & 10.4\%/173 \\
CHA$\rightarrow$0-Container-CFA & 10.5\%/174 & 11.3\%/160 & 10.2\%/185 & 11.6\%/144 \\
CHA$\rightarrow$0-1-CFA  & 10.1\%/184 & 11.3\%/160 & 10.4\%/177 & 10.2\%/177 \\
\midrule
RTA$\rightarrow$0-CFA    & 11.9\%/138 & 11.0\%/141 & 10.2\%/185 & 10.4\%/173 \\
RTA$\rightarrow$0-Container-CFA & 10.5\%/174 & 11.3\%/160 & 10.2\%/185 & 11.6\%/144 \\
RTA$\rightarrow$0-1-CFA  & 10.1\%/184 & 11.3\%/160 & 10.4\%/177 & 10.2\%/177 \\
\bottomrule
\end{tabular}
\end{table}

\subsubsection{Equivalent Algorithm Comparisons}
 Our results reveal a stark divergence in how frameworks interpret modern Java features even though they implemented the same analysis algorithms. Table~\ref{tab:equivalenalg} summarizes the cross-framework consensus. Where possible, we disabled distinct reflection-handling features (e.g., disabling Soot's \texttt{Tamiflex}) to isolate the core graph construction algorithms. Comparisons between theoretically equivalent algorithms in Soot and SootUp  yielded an average similarity of \textbf{32.5\%}. While the tools achieved \textbf{100\% agreement} on features such as Serialization and basic Reflection, they diverged significantly on dynamic benchmarks involving Lambdas and Method References. This confirms that while they share the Jimple IR, their graph construction logic for reflection and \texttt{invokedynamic} differs fundamentally. The differences between Soot and WALA are due to their native model design.

\begin{table}[h]
\caption{Inter-framework for Equivalent Algorithms}
\label{tab:equivalenalg}
\centering
\begin{tabular}{l c c}
\toprule
\textbf{Alg. Pair} & \textbf{Violation} & \textbf{Similarity} \\
\midrule
Soot $\to$ SootUp (CHA) & 240 & 33.2\% \\
Soot $\to$ SootUp (RTA) & 416 & 31.7\% \\
Soot $\to$ WALA (RTA) (BS) & 468 & 21.1\% \\
Soot $\to$ WALA (RTA-FS)& 737 & 20.7\% \\
Soot $\to$ WALA (RTA (OS))& 454 & 21.2\% \\
Soot $\to$ WALA (RTA (OS+FS))& 454 & 21.2\% \\
SootUp $\to$ WALA (RTA (BS)) & 198 & 20.8\% \\
\bottomrule
\end{tabular}
\end{table}

\section{Theoretical Limitations and Insights}
\label{sec:theoretical_limitations}
This section synthesizes the empirical findings from our evaluation to examine the deeper theoretical limitations of static analysis frameworks. Rather than treating the observed violations as isolated defects, we analyze them as symptoms of structural tensions between theoretical assumptions, implementation choices, and practical constraints. Our analysis reveals three fundamental limits: a concentration of semantic fault lines, irreconcilable differences in program construction across frameworks, and a hard ceiling on usable precision. 
\subsection{Distribution of Root Causes}

\label{sec:rootcause_synthesis}
Although the previous sections analyzed intra-framework and inter-framework violations in detail, a global examination of discrepancy logs reveals that semantic inconsistencies are highly skewed rather than uniformly distributed. Across all frameworks, algorithms, and configurations, a small number of semantic mechanisms account for most observed violations.

Static initialization accounts for the largest share of violations, approximately 45\%. Soot eagerly includes \texttt{<clinit>} methods by default, whereas WALA often omits them, creating a persistent baseline discrepancy across programs. Dynamic invocation and lambda-related behavior contribute roughly 30\% of violations and exhibit extreme skew: most benchmark programs remain stable, while a small number of lambda-heavy programs dominate the violation count. In cross-framework comparisons, inputs using \texttt{invokedynamic} triggered more than 1,400 edge mismatches, primarily because WALA resolves bootstrap methods that Soot does not. Reflection-related discrepancies contribute about 15\% of violations. Soot’s default configurations are more conservative, retaining reflective edges that WALA prunes. The remaining violations arise from disagreements over implicit lifecycle methods, such as constructors and finalizers, particularly in virtual-call-heavy programs.
\subsection{The Divergent Construction Problem}
A common assumption in comparative static analysis is that different tools analyze the same logical program when given identical bytecode. Our results   contradict this assumption. We observe that the definition of the analysis scope, including entry points, lifecycle methods, and native boundaries, is framework-specific, creating a semantic gap that no algorithmic refinement can bridge.

Frameworks diverge in their modeling of implicit behavior. For example, Doop treats interactions with java.lang.ref.Finalizer as unreachable noise, while Soot and WALA model these interactions as explicit entry points, producing systematic divergence of approximately 188 edges per program. Frameworks also differ fundamentally in their treatment of native code. Imperative frameworks such as Soot and WALA rely on finite stubs that bound native behavior, whereas Doop propagates data flow through native methods using logic rules. This difference leads to an irreducible precision gap, with Doop detecting over 14,000 additional edges that are mathematically invisible to bytecode-only analysis. These discrepancies are further amplified by modern Java language features. In our cross framework evaluation, SootUp and WALA produced completely disjoint call graphs for the LAMBDA1-3 suites. This indicates not merely differences in precision but mutually incompatible interpretations of the same bytecode. We also observe failures of the closed-world assumption in the presence of reflection. Under object-sensitive configurations, WALA adds edges relative to the baseline (e.g., in NVC4-5), contradicting the expectation that increased precision should strictly prune results. This behavior suggests speculative recovery of reflective targets, highlighting a fundamental tension between soundness, precision, and closed-world reasoning in analysis.

\subsection{The Practical Ceiling of Precision}
Our evaluation also exposes a hard empirical limit on usable precision. Although higher precision analyses are theoretically appealing, they frequently become computationally infeasible on realistic Java workloads. In Doop, the 1-Object-Sensitive analysis repeatedly fails to complete due to solver exhaustion. Even when execution succeeds, intermediate state growth becomes prohibitive. For instance, in the CFNE1 suites, intermediate relation files expand to 133.4 MB. These results suggest that, beyond a certain threshold, precision exists only as a theoretical construct. The refinement lattice remains mathematically valid, but higher nodes in the lattice are empirically unreachable. Under such conditions, expected partial order relationships cannot be observed or validated in practice. This creates a second source of semantic fragility. Violations may arise not from incorrect modeling but from the collapse of tractability at higher precision levels.

\section{Threats to Validity}
\label{sec:validity}

\parhead{Internal Validity.} 
A primary threat is the non-deterministic nature of framework output. We mitigated this via a robust normalization pipeline that resolves internal IDs to semantic signatures. While this rectified format mismatches (raising Soot vs. SootUp similarity from 0\% to 33\%), significant divergence persists, confirming that the remaining gaps are semantic rather than syntactic. 

\parhead{External Validity.} 
We evaluated our approach on the JCG benchmark suite. While these synthetic tests isolate specific language features, they may not reflect the scale of production-grade Java applications. Consequently, the high stability observed in Soot ($100\%$ configuration invariance) might degrade in real-world scenarios with complex reflection usage. Future work could evaluate additional real-world Java programs to further assess generalization.

\parhead{Construct Validity.} 
The construction of call graphs proved to be unstandardized. The inclusion of implicit JVM lifecycle edges by Soot/WALA and their exclusion by Doop created a semantic mismatch of $\approx 188$ edges per program. The 600s timeout on Doop also defined a ``Tractability Wall'' for Object-Sensitive analysis, limiting high-precision comparisons.

\section{Conclusion}
\label{sec:conclusion}

This paper introduces a practical, ground-truth-free approach for detecting semantic violations in Java static analysis frameworks. By leveraging expected partial orders as semantic contracts, we enable systematic testing of analysis algorithms, configurations, and their interactions within/across frameworks. Our evaluation across four popular frameworks reveals that semantic violations are common, subtle, and often silent. We show that higher precision does not reliably imply pruning, that configuration changes can dramatically amplify algorithm-level inconsistencies, and that cross-framework comparisons expose fundamental semantic divergence rather than incidental implementation noise. These findings challenge the assumption that static analysis results are stable or comparable across frameworks. We conclude that semantic validation must explicitly account for algorithm–configuration interactions, and that partial-order-based metamorphic testing provides an effective foundation for building more trustworthy static analysis infrastructure.

\newpage
\bibliographystyle{ACM-Reference-Format}
\bibliography{sample-base}

\appendix

\end{document}